\documentclass[fleqn,twoside]{article}
\usepackage{espcrc2}
\usepackage{epsfig}
\usepackage{graphicx}
\usepackage[figuresright]{rotating}

\newcommand{\AmS}{{\protect\the\textfont2
  A\kern-.1667em\lower.5ex\hbox{M}\kern-.125emS}}

\begin{document}
\title{Study of spectral moments in semileptonic decays of the b hadron 
with the DELPHI detector at LEP.\\
{\small\it Presented at ICHEP02, 
$31^{th}$ Conference on High Energy Physics Amsterdam, July 2002 }
}
\author{M. Calvi
\address{ Dep. of Physics, University of Milano 
Bicocca and INFN, piazza della Scienza 3, I-20126 Milano, Italy}
on behalf of the DELPHI Collaboration
}
\begin{abstract}
The measurement of the moments of hadronic mass spectrum and of 
lepton energy spectrum based on a sample of semileptonic decays of the 
b hadron selected  from $Z^0 \rightarrow b \bar b$  events recorded 
with the {\sc Delphi} detector at LEP, are presented. 
These results are interpreted in terms of constraints on the quark masses 
and on the $b$-quark kinetic energy value.
\vspace{-.8cm}
\vspace{1pc}
\end{abstract}

\maketitle
\section{Introduction}
Moments of hadronic mass spectrum and of lepton energy spectrum are sensitive 
to the masses of the $b$ and $c$ quarks as well as to the non-perturbative 
parameters of the Heavy Quark Expansion.
They allow to improve the determination of the $|V_{cb}|$ element of the 
CKM quark coupling matrix which can be measured from inclusive semileptonic 
B decays by using the relation:
\begin{center}
$\Gamma(b\rightarrow c \ell\nu) = |V_{cb}|^2 f(par.)=
{BR(b\rightarrow c \ell\nu)}/{\tau_b}$
\end{center}
\par\noindent
The current experimental accuracy on the  semileptonic branching ratio 
and B lifetime is about 1\%, while the evaluation of the function $f$, 
based on Operation Product Expansion, brings an uncertainty of the order of 
several percent, making it the dominant error contribution.

Moreover, the comparison of these results with different  measurements 
of the same parameters provides a test of the  consistency of the
theoretical predictions for inclusive semileptonic  B~decays and of 
the underlying assumptions.

Previous measurements of spectral moments have been performed at the 
$\Upsilon(4S)$ \cite{cleo_mom}, 
{\sc Delphi} has performed the first measurement in $b$ hadron semileptonic decays at the $Z^0$.
There are several advantages in the $Z^0$ kinematics, mainly the large boost
 acquired by the b quark ($E_B\sim30$ GeV) which gives access in the 
laboratory frame to the low region of the lepton energy spectrum.
The challenge in this case is the complete reconstruction of the B system.
The lepton energy acceptance extending down  
to the lower end of the spectrum makes these results both 
easier to interpret and complementary to those obtained at the $\Upsilon(4S)$.

In this study two different formulations, using different mass definitions for 
deriving the constraints on the OPE parameters, are used.

\section{Moments of hadronic mass distribution in $b$ hadron semileptonic 
decays}

The hadronic mass distribution of 
${\rm \bar{B^0_d}}\rightarrow {\rm D}^{**}\ell\bar{\nu}$ events have been
studied~\cite{DELPHI_had}.  
D$^{**}$ events have been  reconstructed in the three channels:
D$^0 \pi^+$, D$^+\pi^-$, D$^{*+}\pi^-$  
with the D$^0$, D$^+$ and D$^{*+}$ mesons fully reconstructed.
Leptons have been required to have a momentum greater than 2~GeV/$c$ in the
laboratory frame. 
The separation of the signal from the background has been achieved by means
of a discriminant variable based on the topological properties of the 
secondary vertex such as the presence of additional charged particles in 
addition to the D, the lepton and the neutrino.

D$\pi$ candidates have been separated in ``right sign'' and ``wrong sign'',
considering the charge correlation between the D and the pion, and the 
discriminant variable distributions have been fitted to the simulation
expectation  in the two samples separately.
Only D$^{(*)} \pi^+\pi^-$ events, with one missing pion, can contribute  
to ``wrong sign'' candidates and no evidence of signal have been found 
in this sample. 
The following  upper limits have been derived at 90\% C.L.:

\noindent
$BR(b\rightarrow {\rm D}^0\pi^+\pi^-\ell^-\bar{\nu})=
BR(b\rightarrow {\rm D}^+\pi^+\pi^-\ell^-\bar{\nu})$ 

\noindent
$<0.18\%$
and $BR(b\rightarrow {\rm D}^{*+}\pi^+\pi^-\ell^-\bar{\nu})<0.17\%$.

\begin{figure}
\vspace*{-0.5cm}
\begin{center}
\epsfig{file=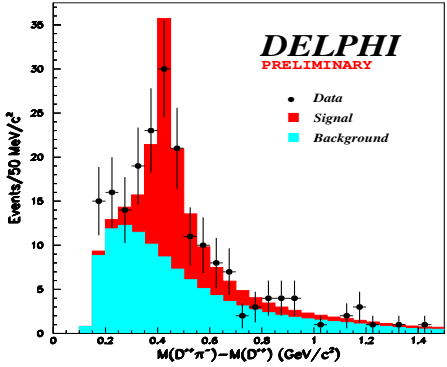,width=6.5cm,height=5.5cm} 
\vspace*{-1.cm}
\caption[]{\sl $\Delta M$ distributions in the D$^{*+}\pi^-$ 
reconstructed decay channel.}
\end{center}
\label{fig:deltam}
\vspace*{-0.8cm}
\end{figure}

Therefore only D$\pi$ states have been considered in the following 
analysis of the hadronic mass distribution of the D$^{**}$ states.
A fit to the variable $\Delta M=m({\rm D}^{(*)}\pi)-m({\rm D}^{(*)})$ 
has been performed,
considering the contributions of narrow and broad resonant states  
D$_0^{*+}$, D$_1^{*+}$, D$_1^{+}$ and D$_2^{+}$  as well as non resonant 
D$\pi$ states. An example of $\Delta M$ distribution is shown in 
%Figure~\ref{fig:deltam}.
Figure~1.
%For completely reconstructed D$\pi$ states a resolution of 4 MeV have been 
%obtained, with 4 and 15 MeV of additional smearing due to missing $\pi^0$ 
%or photons, respectively.

The total rate for D$^{**}$ production amounts to:
\par\noindent
${\rm \bar{B^0_d}}\rightarrow {\rm D}^{**}\ell\bar{\nu}=(2.6\pm0.5\pm0.6)\%$ 
and the broad D$_1^{*+}$ is the dominant contributing channel.

Moments of the D$^{**}$ mass distribution have been evaluated from the 
fitted mass distributions. To determine the moments of the complete 
hadronic mass distribution, in $b$-hadron semileptonic decays, the 
$b\rightarrow {\rm D}\ell\bar{\nu}$ and  
$b\rightarrow {\rm D}^*\ell\bar{\nu}$ components have been included using 
the relation
$ \langle m_H^n \rangle = p_D m_D^n +  p_{D^*} m_{D^*}^n +
                        p_{D^{**}} \langle m_{D^{**}}^n \rangle $
where $p_{D}$ and $p_{D^*}$ are the relative branching fractions derived  
from published results and $p_{D^{**}}$ is obtained by imposing the 
constraint $p_D +  p_{D^*} + p_{D^{**}} =1$ and using the above measurement.

The following preliminary DELPHI results have been obtained:
\begin{center}
$\langle m_H^2 - m_{\bar D}^2 \rangle = (0.534 \pm 0.041
                                          \pm 0.074 ) ~GeV/c^2$\\
$\langle (m_H^2 - m_{\bar D}^2)^2 \rangle = (1.51 \pm 0.20
                                          \pm 0.23 ) ~(GeV/c^2)^2$\\
$\langle (m_H^2 - \langle m_H^2\rangle )^2 \rangle = (1.23 \pm 
                                                0.16 \pm 0.15 ) ~(GeV/c^2)^2$\\
$\langle (m_H^2 - \langle m_H^2\rangle )^3 \rangle = (2.97 \pm 
                                                0.67 \pm 0.48 ) ~(GeV/c^2)^3$\\
\end{center}
\par\noindent
where the first uncertainty is statistic and the second is systematic.

\section{Moments of  lepton energy spectrum  in $b$ hadron semileptonic 
decays}

The moments of the lepton energy spectrum provide constraints similar to 
those of the hadronic mass. Further they offer an important consistency 
test of the theory.

An inclusive reconstruction of the semileptonic decays has been 
performed~\cite{DELPHI_lept}.   
Muons and electrons with a momentum greater than 2.5 and 3.0 GeV/c,
 respectively, have been tagged in a sample of $Z^0\rightarrow b \bar{b}$ 
events. Secondary vertices have been reconstructed using an 
iterative procedure. The B energy has been reconstructed adding to the 
 charm vertex energy, the lepton energy and the neutrino energy, 
evaluated from the event missing energy. 
\begin{figure}[hb]
\vspace*{-0.5cm}
\begin{center}
\epsfig{file=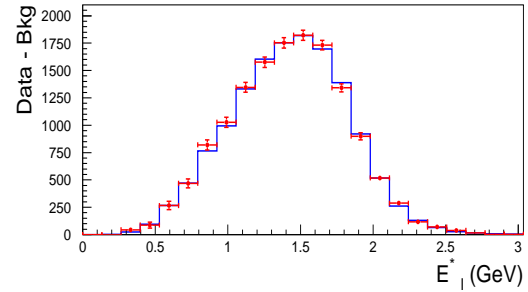,width=7.cm,height=4.0cm}
\vspace*{-1.cm}
\caption[]{\sl Lepton energy spectrum after background subtraction. 
Points are data and histogram is the result of the fit.}
\end{center}
\label{fig:lep_spectrum}
\vspace*{-1.cm}
\end{figure}
The B direction has been estimated from both the 
reconstructed B momentum and the B decay flight direction. 
Boosting the leptons in the B reconstructed rest frame provides a 
typical energy resolution of 250 MeV.
In order to separate signal $b \rightarrow \ell$ events form 
$b \rightarrow c \rightarrow \ell$ and other backgrounds without introducing
significant biases to the lepton energy distribution, two probabilistic 
variables have been defined based on charge correlation and event topology.
The measured energy spectrum after background subtraction 
is shown in 
%Figure~\ref{fig:lep_spectrum}. 
Figure~2.
For this preliminary result only 1994 and 1995 statistics have been used, 
corresponding to a sample of 18,300 leptons.
After unfolding the resolution smearing, the first, second and third moments 
have been calculated. 
The relevant corrections for distortions due to electromagnetic radiation, 
contamination of $b \rightarrow u \ell \nu$ decays and
contribution of B$_s^0$ and $\Lambda_b$ hadrons in the events sample
have been applied.
The following preliminary DELPHI results have been obtained:
\begin{center}
$\langle E_\ell^* \rangle = (1.383 \pm 0.012 \pm 0.008 )~GeV$\\
$\langle (E_\ell^*-\langle E_\ell^*\rangle)^2\rangle = (0.192 \pm 0.005 
\pm 0.010 )~GeV^2$\\
$\langle (E_\ell^*-\langle E_\ell^*\rangle)^3\rangle = (-0.029 \pm 0.005 
\pm 0.005 )~GeV^3$\\
\end{center}
where the first uncertainty is statistic and the second is systematic.

\section{Interpretation of the results}

Two different approaches have been followed in order to obtain constraints 
on the non-perturbative parameters of OPE from the measured spectral moments.
The first~\cite{Falk} is based on an expansion on the pole masses
 $m_b$ and $m_c$ and expresses the $b$-energy parameter as $\lambda_1$, 
while the second ~\cite{Voloshin} uses running heavy quark masses 
$m_b(\mu)$ and $m_c(\mu)$ and the kinetic energy expectation value 
$\mu_{\pi}^2$, corresponding to $\lambda_1$. 

Results obtained from the measured values of the first two moments of the 
hadronic mass spectrum and lepton energy spectrum have been found to be 
compatible. An exemplification is given in 
%Figure~\ref{fig:lambda_Lambda} 
Figure~3
showing constraints extracted in the $\bar{\Lambda}$ - $\lambda_1$ 
plane which give:
\begin{center}
$\bar{\Lambda}=(~0.44 \pm 0.04 \pm 0.05 \pm0.07 )~GeV$\\
$\lambda_1 =(-0.23 \pm 0.04 \pm0.05 \pm0.08 )~GeV^2$
\end{center}
\par\noindent
where the quoted uncertainties are statistic, systematic and
related to the power corrections and $\alpha_s$ uncertainties, respectively. 
\begin{figure}[h!]
\vspace*{-0.6cm}
\begin{center}
\epsfig{file=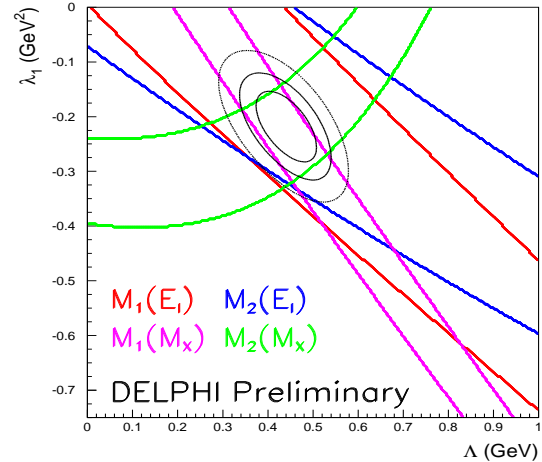,width=8.0cm,height=7.0cm} 
\vspace*{-1.6cm}
\caption[]{\sl The constraints on the $\bar{\Lambda}$ - $\lambda_1$ plane 
obtained from the combination of the first two moments of the lepton energy 
$(M_1(E_\ell)$,$M_2(E_\ell))$ and hadronic mass $(M_1(M_X)$,$M_2(M_X))$.
The bands represents the regions selected by each 
moment within $\pm 1 \sigma_{stat+syst}$. The ellipsis show the $1 \sigma$, 
68\% and 90\% C.L. respectively.}
\end{center}
\label{fig:lambda_Lambda} 
\vspace*{-1.0cm}
\end{figure}

Another way to express these results exploits the correlation in 
the expected values for the charm and beauty quark masses 
to extract the charm quark mass $m_c$ 
by using an independently determined value of $m_b$. 
Using the first two moments of the lepton energy spectrum and 
$m_b$=(4.60$\pm$0.05 ) GeV, with the second theoretical formulation we find: 
$$m_c{\mathrm{(1~GeV)}} = (1.19 \pm 0.05 \pm 0.05 \pm 0.07  \pm 0.04 )~GeV$$
 where the quoted uncertainties are statistic, systematic
$\pm$~0.050~GeV on $m_b$ and $1/m_b^3$ corrections, respectively. 

\section{Acknowledgment}

I would like to thank M.Battaglia and P.Roudeau for their contributions 
to this work, N.Uraltsev and P.Gambino for contributing
to the theoretical predictions used in this paper.


\begin{thebibliography}{99}

\small

\bibitem{cleo_mom}
J.~Bartelt {\it et al.} (CLEO Collaboration), CLEO-CONF 98-21.
D.~Cronin-Hennessy {\it et al.} (CLEO Collaboration), Phys. Rev. Lett.
{\bf 87} (2001)251808

\bibitem{DELPHI_had} 
D.~Bloch {\it et al.}, DELPHI~2002-070~CONF~604, 
updated from contributed paper to 
%ICHEP02, $31^{th}$ Conference on 
%High Energy Physics Amsterdam, July 2002.
contributed paper to this Conference.
\bibitem{DELPHI_lept}
M.~Battaglia {\it et al.}, DELPHI 2002-071 CONF~605, 
%{\it A Study of the Lepton Spectrum Moments in 
%$b \rightarrow  X_c \ell \bar\nu$ Decays with the DELPHI Detector at LEP}
contributed paper to this Conference.
%ICHEP02, $31^{th}$ Conference on 
%High Energy Physics Amsterdam, July 2002.

\bibitem{Falk} A.~F.~Falk and M.~Luke, Phys. Rev. {\bf D57} (1998)424.
P.~Gambino, private communication.

\bibitem{Voloshin} M.B.Voloshin Phys. Rev. {\bf D51} (1995)4934.\\
I.~I.~Bigi, M.~A.~Shifman and N.~Uraltsev Ann. Rev Nucl.Part. Sci {\bf47} 
(1997)591. N.~Uraltsev, private communication.

\end{thebibliography}
\end{document}